# Giant magnetostriction in La$_2$CoMnO$_6$ synthesized by microwave irradiation

M. Manikandan, A. Ghosh and R. Mahendiran[*]

*Department of Physics, National University of Singapore, 2 Science Drive 3,*

*Singapore 117551, Republic of Singapore*

## ABSTRACT

Polycrystalline insulating ferromagnetic double perovskite La$_2$CoMnO$_6$ possessing monoclinic structure and a high ferromagnetic Curie temperature ($T_C$ = 222 K) was rapidly synthesized (~30 min) by irradiating stoichiometric mixture of oxides with the microwave. The sample exhibits negative magnetostriction ($\lambda_{\text{par}}$), i.e., contraction of length along the magnetic field direction in the ferromagnetic state. At 10 K, $\lambda_{\text{par}}$ does not show saturation up to a magnetic field of 50 kOe where it reaches 610 x 10$^{-6}$ which is one of the highest values of magnetostriction found so far among perovskite oxides with 3$d$ ions. The magnitude of $\lambda_{\text{par}}$ decreases monotonically as the temperature increases and becomes negligible above $T_C$. The giant magnetostriction in this double perovskite is suggested to originate from large spin-orbit coupling associated with Co$^{2+}$ ($d^7$) cation. The obtained magnetostriction value is comparable to $\lambda_{\text{par}}$ = 630 x 10$^{-6}$ in an identical composition obtained through solid-state reaction over several days in a conventional furnace which indicates the advantages of microwave-assisted synthesis in saving reaction time and electric power without deteriorating physical properties.

[*]Author for correspondence: R. Mahendiran (email: *phyrm@nus.edu.sg*)

While the low-temperature ground state of the perovskite oxide LaBO$_3$ is a non-magnetic insulator (antiferromagnetic insulator) for B = Co$^{3+}$ (Mn$^{3+}$) cations, LaCo$_{0.5}$Mn$_{0.5}$O$_3$ is a ferromagnetic insulator with a high Curie temperature ($T_C$ = 220-225 K).[1] G. Blasse[2] was the first to suggest that Co and Mn ions adopt di- and tetra- valent states, respectively in this composition and proposed that they order next to each other in rock salt configuration with negatively charged oxygen intervening them. Hence, La$_2$CoMnO$_6$ is a double perovskite with twice the lattice spacing of LaCo$_{0.5}$Mn$_{0.5}$O$_3$. Blasse argued that ferromagnetism in this compound originates from positive superexchange interaction among high-spin Co$^{2+}$ ($t_{2g}^5 e_g^2$, $S$ = 3/2) and high-spin Mn$^{4+}$ ($t_{2g}^3 e_g^0$, $S$ = 3/2) ions rather than double exchange interaction between Mn$^{3+}$-Mn$^{4+}$ pairs.[3] The presence of Co$^{2+}$ and Mn$^{4+}$ cations in La$_2$CoMnO$_6$ was confirmed by different researchers using X-ray near-field edge absorption and magnetic dichroism.[4,5,6,7] Excess oxygen non-stoichiometry or excess deviation from the equimolar proportion of Co$^{2+}$ and Mn$^{4+}$ lowers the Curie temperature ($T_C$ = 80-180 K) and/or induces double magnetic transitions due to the creation of Mn$^{3+}$ and Co$^{3+}$ ions besides Co$^{2+}$ cation.[8,9,10,11,12]

Recent resurge of interest in La$_2$CoMnO$_6$ is due to the frantic search for multiferroism, i.e., the coexistence of spontaneous electrical and magnetic dipoles ordering, and magnetoelectric coupling between them. However, magnetoelectric coupling in La$_2$CoMnO$_6$ seems to be weak based on the analysis of magnetocapacitance data.[13,14,15] Although Raman spectroscopy indicates non-negligible spin-phonon interaction,[16,17] there is no convincing evidence for the presence of ferroelectricity in this ferromagnetic insulating oxide until now. On the other hand, the presence of Co$^{2+}$ ion with an unquenched orbital moment in this double perovskite will be interesting from the viewpoint of magnetostriction, i.e., magnetic-field induced strain which is quantified as the fractional change in length *λ* = [*L(H)*-*L(0)*]/*L(0)*, where *L(H)* and *L(0)* are the lengths of the sample in a magnetic field *H* and *H* = 0 at a constant temperature. Among oxides, CoFe$_2$O$_4$ possessing divalent cobalt exhibits the highest magnetostriction at room temperature (*λ*$_{100}$ = - 590 x 10$^{-6}$ in Co$_{0.8}$Fe$_{2.2}$O$_4$ single crystal[18] but its magnitude decreases to ~ 120-400 x 10$^{-6}$ in polycrystals depending on the synthesis conditions[19]) but it has spinel structure and the magnitude of magnetostriction decreases below 200 K due to rapid increase in magnetic ansiotropy. Despite extensive studies on La$_2$CoMnO$_6$, magnetostriction has not been reported so far and it is the motivation of this report.

In this work, we synthesized polycrystalline $La_2CoMnO_6$ samples by two methods for comparison: 1. Exposing oxide precursors to microwave (MW) irradiation for a short time (~10-30 min), and 2. Solid-state reaction method over several days using a conventional electrical furnace. Microwave-assisted synthesis is a novel and emerging method of synthesizing organic and inorganic materials but its potential is not fully exploited yet.[20,21,22] While a sample placed inside a conventional electric furnace is heated from surface to interior by transfer of heat from a high current carrying SiC filaments to the container wall of the sample, microwave energy is directly converted into heat in a microwave furnace through the interaction of microwave electromagnetic field with polar molecules in the sample, and heat generated in the sample diffuses from interior to surface of the sample. Heat is generated mainly due to frictional forces when electrical dipoles in the material are unable to follow the oscillating electric field component of the impinging MW radiation. Besides electrical dipole oscillations, migration of ions, eddy current and magnetic hysteresis losses can also contribute to MW heating although the major contribution is by dielectric heating caused by oscillating dipoles. However, microwave heating requires at least one of the precursors to make the required sample should be a good microwave absorber, i.e., it should have a high dielectric loss in the allowed operating microwave frequency (2.45 GHz). The precursor which absorbs MW power acts like a hot spot and ignites a reaction among other precursors. While it takes about several days to obtain a single-phase pellet using a conventional furnace, it takes only 1 to 45 min to synthesize a single-phase pellet by microwave irradiation depending on the microwave absorbing property of precursors. It is of fundamental interest to know how ultra-fast heating in MW-assisted synthesis influences cation ordering and subsequently affects the magnetic and other properties. In earlier work, $La_2CoMnO_6$ nanoparticles were synthesized by co-precipitation method involving nitrates of La, Co and Mn and the powder was sintered in a domestic MW-furnace.[23] In the present work, the chemical reaction among oxide precursors was initiated by MW irradiation and the pellet was also sintered in the MW furnace.

We prepared 10 grams of $La_2CoMnO_6$ powder from the stoichiometric ratio of dehydrated $La_2O_3$, $Co_3O_4$ and $Mn_2O_3$ by mixing and grinding them using an agate mortar and pestle. Five grams of the mixed powder taken in an alumina crucible was placed at the centre of a muffle microwave furnace (Milestone PYRO, model MA 194-003) operating at 2.45 GHz and irradiated the powder with microwave (MW) power of 1600 W for 10 min. The temperature profile was set to reach 1000 °C in 10 min (heating rate of 100 °C/min) where it was maintained for 20 min. Then, the MW power was switched off and the powder cooled to room temperature naturally (~90 min). The calcined powder was ground again, pelletized, and irradiated with a microwave to reach 1200 °C in 10 min and dwelled at that temperature for 20 min before switching off the power. The sample cools to room temperature in 4 hr (12 °C/min). Another five grams of powder was taken in an alumina crucible and calcined in the conventional resistive heating furnace at 1000 °C for 12 hours, followed by grinding and calcining at 1100 °C for another 12 hours, followed by pelletizing and sintering the pellet at 1200 °C for 12 hours. The heating and cooling rates were 5 °C/min. The obtained samples were characterized using powder X-ray diffraction Cu-K$_{\alpha 1}$ (1.5406 Å) for purity and structure. Magnetization was measured using a vibrating sample probe for a physical property measurement system (PPMS). Magnetostriction along the direction of the applied *dc* magnetic field ($\lambda_{par}$) was measured using a capacitance dilatometer probe inserted in the PPMS in the temperature range from 300 K to 10 K.[24] A polished cubic sample of size 2 x 2 x 2 mm was sandwiched between two circular capacitive electrodes in the dilatometer probe. The change in the capacitance of the dilatometer was measured using a high-resolution capacitance bridge (Andeen Hagerling, model AH2500A).

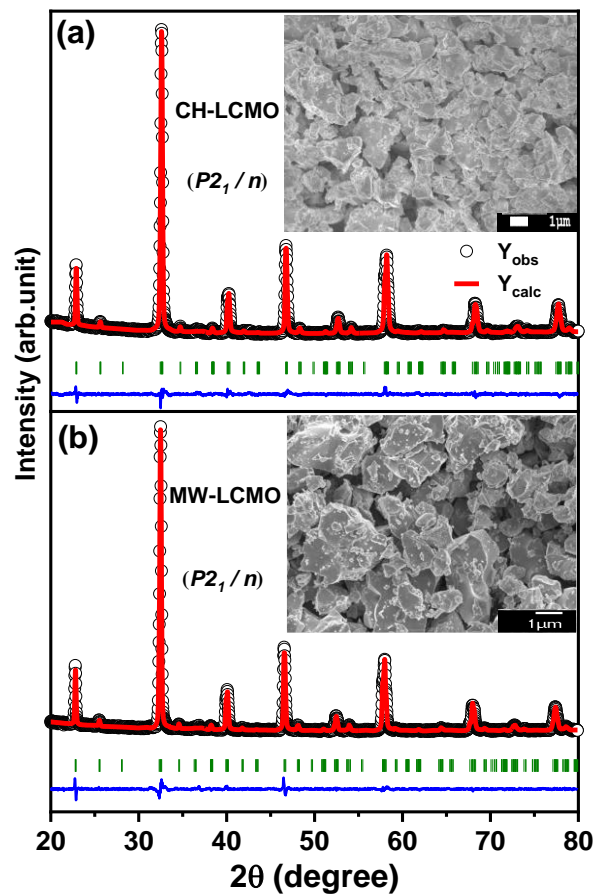

**Fig. 1**. Powder X-ray diffraction pattern (symbol) and Rietveld fit for La$_2$CoMnO$_6$ synthesized by **(a)** conventional heating (CH-LCMO) in an electrical furnace and **(b)** microwave irradiation (MW-LCMO). The inset in each plot shows the scanning electron microscope image.

Fig. 1(a) and (b) compare the X-ray diffraction patterns of the La$_2$CoMnO$_6$ samples synthesised via (a) microwave (MW) irradiation method and (b) solid-state reaction route using conventional heating (CH) in an electrical furnace. Hereafter, we refer them as MW-LCMO and CH-LCMO samples, respectively. Both the samples are found to be single-phase and crystallized in a monoclinic structure (*P2$_1$/n* space group) at room temperature. The refined lattice parameters are *a* = 5.534(1) Å, *b* = 5.501(3) Å and *c* = 7.788(3) Å, *β* = 89.96(11) and *V* = 237 Å$^3$ for the MW-LCMO and *a* = 5.5137(1), *b* = 5.4798(4), *c*= 7.7601(6), β = 89.90(7), V = 234 Å$^3$ for the CH-LCMO, which are in close agreements with previous reports.[7,8] The insets in both figures show scanning electron microscopy images of the respective samples. While both the samples show heterogeneous shape and size distribution, relatively bigger particles are seen in MW-LCMO.

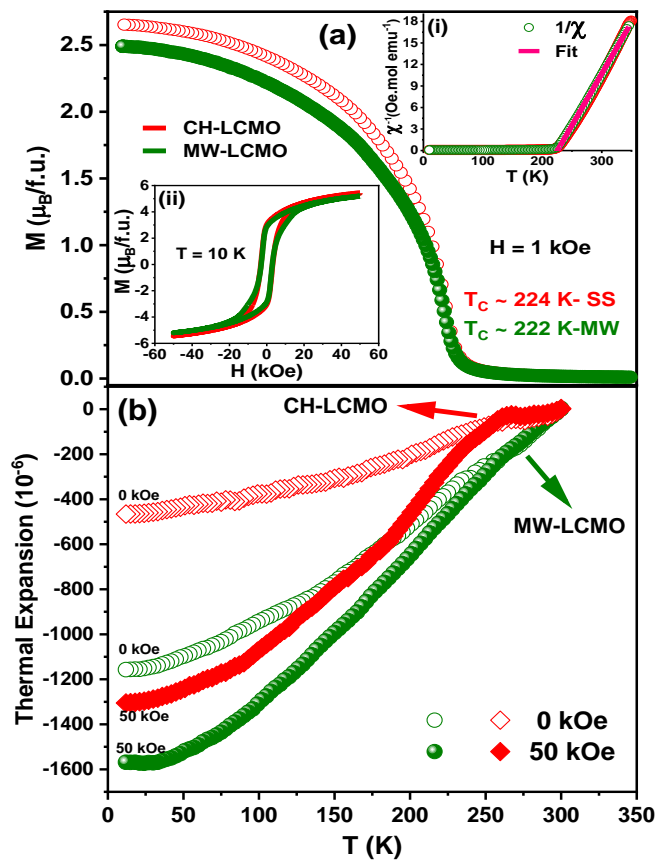

**Fig. 2**. **(a)** Main panel: Temperature dependence of magnetization $M(T)$ measured in $H = 1$ kOe for MW-LCMO (closed circle) and CH-LCMO (open circle). Top inset: Temperature dependence of inverse susceptibility for MW-LCMO and Curie-Weiss fit. Bottom inset: *M-H* isotherms at 10 K in both samples. **(b)** Temperature dependence of thermal expansion under $H = 0$ and 50 kOe magnetic fields for MW-LCMO (circle) and CH-LCMO (square).

The main panel of Fig. 2(a) shows the temperature dependence of magnetization (*M*) in both MW-LCMO and CH-LCMO samples for $H = 1$ kOe upon cooling from 350 K to 10 K. The rapid increase of $M(T)$ around 230 K signals the onset of ferromagnetic transition. The ferromagnetic Curie temperature ($T_C$) estimated from the inflexion point of d$M$/d$T$ curve is $T_C$ = 224 K (222 K) for the CH-LCMO (MW-LCMO). At $H = 1$ kOe, magnetization of the MW-LCMO sample is slightly lower than the CH-LCMO. The right top inset depicts the temperature dependence of inverse susceptibility of the MW sample which increases linearly with the temperature above $T_C$. From the Curie-Weiss fit ($1/\chi = (T-\theta_W)/C$) in the paramagnetic state, we obtain a positive Weiss temperature $\theta_W = 225$ K and an effective magnetic moment $\mu_{eff} = 5.29$ $\mu_B$, which are closer to the values reported by Burnus et al.[4] and Bull et al.[7] Since the inverse susceptibility for the CH sample is identical to the MW sample, we don't show it here. The left bottom inset compares *M-H* isotherms at 10 K in both samples, which indicate ferromagnetic behaviour with close coercive fields ($H_C$ = 3.12 (3.18) kOe for MW (CH) samples). However, *M* does not show saturation in the maximum field $H = 50$ kOe where it attains 5.18 $\mu_B$/f.u. (5.41 $\mu_B$/f.u.) for the MW (CH) sample. These values are lower than 6 $\mu_B$/f.u. expected for ferromagnetically aligned spins of fully ordered $Mn^{4+}$ and $Co^{2+}$ ions[8]. The decrease from the theoretical value is possibly due to the misplacement of some $Co^{2+}$ ions at the $Mn^{4+}$ site or vice versa (antisite disorder) which promotes antiferromagnetic interaction among like-charged neighbouring ions.[25] Fig. 1(b) shows thermal expansion as a function of temperature in both samples for $H = 0$ (open symbols) and 50 kOe (closed symbols). The length of each sample contracts with decreasing temperature as expected for anharmonic vibration of ions but the contraction is relatively larger in the MW-LCMO sample. While no anomaly is visible in linear thermal expansion under zero-field around $T_C$, the thermal expansion under $H = 50$ kOe shows a clear deviation from the zero-field curve starting from $T_C$ down to 10 K in both samples.

Fig. 3(a) shows the field dependence of the parallel magnetostriction ($\lambda_{par}$) at 10 K on the left $y$- scale and squared magnetization plotted as -$M^2$ on the right $y$-axis for MW-LCMO. The length of the sample shrinks as the magnetic field increases, i.e., $\lambda_{par}$ shows the negative magnetostriction effect. We can also see that $\lambda_{par}$ exhibits double peaks in the low-field region that coincide with coercive fields ($H_c$ = 1.8 kOe) in $M(H)$ while reversing the field direction. Hysteresis in -$M^2$ is wider than in $\lambda_{par}$ and it extends to higher field than in magnetostriction. Above the coercive field, $\lambda_{par}$ shrinks smoothly without showing saturation up to the maximum available field. At 50 kOe, $\lambda_{par}$ = -610 x $10^{-6}$ and its magnitude is higher than +500 x $10^{-6}$ found earlier in the ferromagnetic metallic oxide $La_{0.5}Sr_{0.5}CoO_3$ for the same field strength.[26] However, the sign of $\lambda_{par}$ in $La_2CoMnO_6$ is opposite to that of $La_{0.5}Sr_{0.5}CoO_3$. It could be related to the direction of the easy axis and the valance state of Co ions. $La_2CoMnO_6$ is monoclinic whereas $La_{0.5}Sr_{0.5}CoO_3$ is cubic at room temperature and the Co cation exists mixed valent ($3^+/4^+$) in the latter compound. The field dependence of $\lambda_{par}$ at 10 K in the CH-LCMO (Fig. 3(b)) is similar to that of the MW sample and its magnitude at 50 kOe is -630 x $10^{-6}$ is slightly higher than the MW sample.

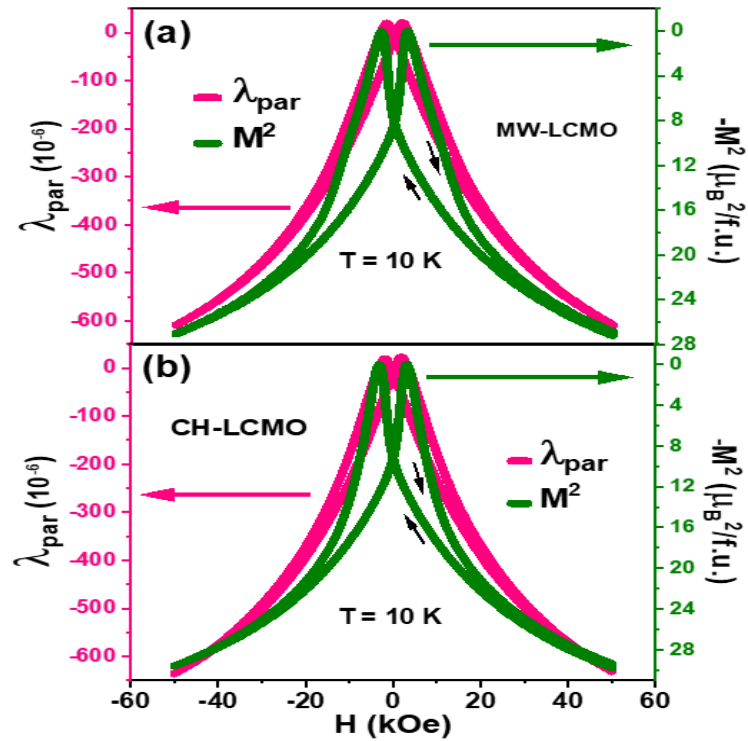

**Fig. 3. (a)** Magnetic field dependence of the parallel magnetostriction $\lambda_{par}$ (left y-axis) and -$M^2$ (right y-axis) at 10 K for the MW-LCMO. The same are shown in **(b)** for the CH-LCMO.

The field dependence of $\lambda_{par}$ at selected temperatures from $T = 25$ K and up to 300 K are shown in the insets of Fig. 4(a) and (b) for the MW and CH samples, respectively, only for the positive fields for simplicity. From these isothermal field sweep data, we extract the $\lambda_{par}$ for $H = 50$, 30 and 10 kOe and plot them in the main panel of (a) and (b) for the MW and CH samples, respectively. The $\lambda_{par}$ for 50 kOe decreases only a little between 10 K and 75 K but more rapidly between 100 K and 225 K, and becomes negligibly small above 225 K in the temperature region where the sample is paramagnetic. A similar trend is also observed for $H = 30$ kOe and 10 kOe and also in the CH-LCMO (fig. 4(b)).

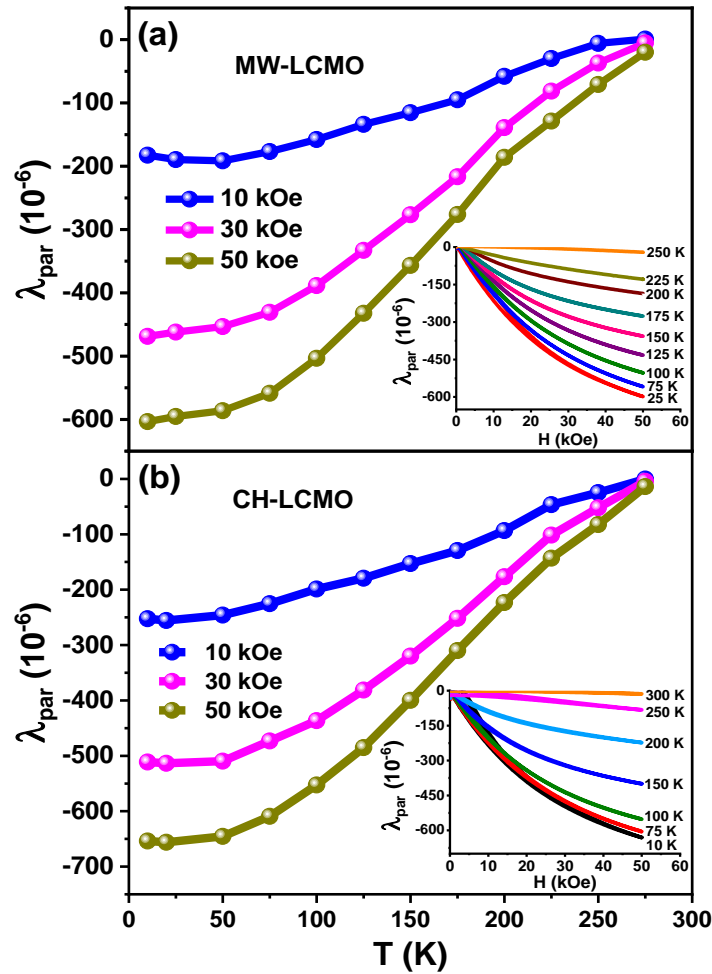

**Fig. 4. (a)** Inset: The field dependence of $\lambda_{par}$ at selected temperatures for MW-LCMO and the main panel shows the temperature dependence of $\lambda_{par}$ for $H = 10$, 30 and 50 kOe extracted from the data shown in the inset. The same quantities are shown in **(b)** for the CH-LCMO.

The observed giant magnetostriction in La$_2$CoMnO$_6$ most likely originates from spin-orbit interaction (single ion mechanism) at the Co$^{2+}$ cation. High-spin Co$^{2+}$:d$^7$ in octahedral coordination has unquenched orbital momentum as encountered in the spinel CoFe$_2$O$_4$.[27] Since intra-atomic exchange energy is large than crystal field energy for Co$^{2+}$, exchange interaction lifts spin degeneracy of 5 fold degenerate 3$d$ level of free Co ion into spin-up and spin-down levels. The cubic crystal effect due to six O$^{2-}$ ions splits the five-fold degenerate 3$d$ states of Co$^{2+}$ cations into triply degenerate $t_{2g}$ and doubly degenerate $e_g$ levels. Spin-up $t_{2g}$ and $e_g$ levels are filled by 3 and 2 spin-up electrons, respectively, the remaining 2 electrons occupy $t_{2g}$-spin down states. The down-spin $t_{2g}$ state is further split into a singlet ($<L_z> = 0$) and a doublet ($<L_z> = 1$) by trigonal crystal field arising from the next neighbour Mn$^{4+}$ cations along [111] axis. The $t_{2g}$ doublet has wave functions that spread into the plane perpendicular to the trigonal axis and hence the angular momentum is parallel to the trigonal axis. X-ray magnetic dichroism experiments in La$_2$CoMnO$_6$ have confirmed that Co$^{2+}$ possess significant orbital magnetic moment while that of Mn$^{4+}$ is negligible in comparison. The ratio of orbital to spin magnetic moment $m_{orb}/m_{spin}$ varies from 0.47 to 0.649 in bulk polycrystalline samples depending on synthesis methods used,[4-5] and $m_{orb}/m_{spin}$ = 0.637 (0.456) in 15 nm thin films of La$_2$Co$_{1-x}$Mn$_{1+x}$O$_6$ ($x = 0.23$) grown on SrTiO$_3$ (LSAT) substrate.[6] Due to strong intra-atomic spin-orbit coupling, the orbital angular moment also rotates along with the spin moment when the latter tend to align along the field direction. This leads to the contraction of the length. As the external magnetic field is increased from zero, a polycrystalline CoFe$_2$O$_4$ sample initially contracts in length in low fields (< 3 kOe) but expands at higher fields since low field behaviour is dominated by negative magnetostriction along [100] axis and positive magnetostriction along [111] direction at higher fields.[19] Also, the origin of magnetostriction in LCMO is different from that of the metallic ferromagnetic perovskite La$_{0.5}$Sr$_{0.5}$CoO$_3$.[26,28] Magnetostriction in the latter compound was attributed to magnetic field-induced low-spin ($S = 0$) to intermediate-spin ($S = 1$) transition of Co$^{3+}$ ions and resulting orbital instability of the Jahn-Teller active intermediate spin Co$^{3+}$. Parallel magnetostriction in ferromagnetic manganites (eg., La$_{1-x}$A$_x$MnO$_3$, A= Sr, Ca; $x = 0.3$-0.33) is an order of magnitude lower ($\approx$ +40 x 10$^{-6}$ at 10 K) than the titled compound.[29] Although observed magnetostriction in polycrystalline LCMO is lower than metallic alloys such as Terfenol-D (Tb$_{1-x}$Dy$_x$Fe$_2$) and Galfenol (Fe$_{1-x}$Ga$_x$ alloy),[30] LCMO is mechanically robust over a wide temperature, easier to fabricate with low-cost raw materials and highly insulating than the above alloys.

In conclusion, the ferromagnetic double perovskite $La_2CoMnO_6$ samples were obtained within 30 min by microwave irradiation and using a conventional electrical furnace that takes several days. However, they showed comparable magnetic and magnetostrictive properties. Both the samples exhibit giant values of magnetostriction (~ 610-630 microstrain at 10 K in a field of 50 kOe), whose origin was suggested to be the spin-orbit coupling of $Co^{2+}$. For future work, it will be interesting to explore the impact of cation disordering on magnetostriction and its angular dependence. Further, the $LaCo_{1-x}Mn_xO_3$ series will be worthy of investigation to understand the dependence of magnetostriction on a shift in the valency states of the transition metal ions with changing $x$.

## ACKNOWLEDGEMENTS

R. M. acknowledges the Ministry of Education, Singapore, for supporting this work (Grant no. A-8000924-00-000)

## AUTHOR DECLARATIONS

**Conflict of Interest**

The authors have no conflicts to disclose.

## DATA AVAILABILITY

The data that support the findings of this study are available from the corresponding author upon reasonable request.